# How does an AI Weather Model Learn to Forecast Extreme Weather?

**Rebecca Baiman[1], Elizabeth A. Barnes[1], and Ankur Mahesh[3,4]**

[1]Department of Atmospheric Science, Colorado State University, Fort Collins, CO, USA.

[3]Earth and Environmental Sciences Area, Lawrence Berkeley National Laboratory (LBNL), Berkeley, CA, USA.

[4]Department of Earth and Planetary Science, University of California, Berkeley, CA, USA

Corresponding author: Rebecca Baiman (rbaiman.earth@gmail.com)

**Key Points:**

- We present the first look at an AI weather model's forecasts of extreme weather events through 90 epochs of training.
- Forecasts of extreme weather can improve then degrade through training as the time-averaged error is minimized.
- To improve forecasts of extreme weather, AI weather models may be able to leverage information learned earlier in training.

## Abstract

In a warming climate with more frequent severe weather, artificial intelligence (AI) weather models have the potential to provide cheaper, faster, and more accurate forecasts of high-impact weather events. To realize this potential, there is a need for more research on how models learn extreme events and how that learning might be improved. We investigate how a spherical Fourier neural operator model (SFNO) learns extreme weather by saving every checkpoint throughout training and analyzing a collection of 9 extreme weather events including heatwaves, atmospheric rivers, and tropical cyclones. The SFNO learns heatwaves similarly to other weather days, but we find evidence that the model learns information about atmospheric river and tropical cyclone forecasts that it loses later in training. We propose a possible training strategy to improve the forecasting of extreme events by retaining information from earlier training checkpoints, and provide initial evidence of its utility.

## Plain Language Summary

A new frontier of weather forecasting uses artificial intelligence (AI) instead of traditional physics-based models to make forecasts. These AI models have the potential to be cheaper, faster, and more accurate than traditional models. However, AI weather models still have room for improvement in forecasting extreme weather events like hurricanes. We look at how one of these machine learning models learns extreme weather during training of the model. We find that for atmospheric rivers and tropical cyclones, the model can produce a better forecast early in it's training, but this information is lost by the end of training. Our results suggest that the development of AI weather models could potentially improve extreme weather forecasts by adjusting training methods.

## 1 Introduction

As a warming climate brings more frequent and more extreme weather events ("IPCC AR6 Working Group 1," 2021), artificial intelligence (AI) emulators open the door to faster, cheaper, and potentially more accurate weather forecasts compared to physics-based models (Pasche et al., 2025). These emulators, including but not limited to Google's GraphCast (Lam et al., 2023), Huawei's PanguWeather (Bi et al., 2023), and NVIDIA's FourCastNetV2 (Bonev et al., 2023), are trained on European Center for Mediumrange Weather Forecasts Reanalysis v5 (ERA5) and use variations of mean squared error or mean absolute error as their loss function. Since skillful forecasting of extreme weather is integral to the utility of these emulators, evaluation based on time-averaged forecast error is inadequate (Olivetti & Messori, 2024a). A growing body of research addresses the skill of these AI emulators in forecasting extreme weather (Charlton-Perez et al., 2024; Olivetti & Messori, 2024a, 2024b; Pasche et al., 2025; Sun, Hassanzadeh, Zand, et al., 2025). One recent study finds that the AI emulators Pangu-Weather and GraphCast outperform state-of-the-art physics-based IFS HRES in forecasting near-surface temperatures and wind on average. However, the best model at forecasting extremes for these variables depends on region, type of extreme, and lead time (Olivetti & Messori, 2024b).

Here, we focus on three types of high-impact events with varying time scales of evolution: heatwaves, atmospheric rivers (ARs), and tropical cyclones (TCs). Previous work finds varied results of how well AI emulators forecast these types of events. For forecasts up to 6.5 days, AI emulators underperform IFS HRES in forecasting the out-of-distribution 2021 PNW Heatwave even though their performance is comparable in forecasting more typical days in the Summer 2022 (Pasche et al., 2025). The intensity and accuracy of individual AR forecasts from AI emulators remains an open question, but AI emulators including FourCastNet V2 underperform IFS HRES in identifying these events for lead times within four days (Davis et al., n.d.). AI emulators capture seven-day TC tracks with comparable skill to operational forecast models, but substantially underestimate the cyclone intensity (DeMaria et al., 2025). Based on this current literature examining heatwaves, ARs, and TCs, AI emulators are not yet comparable to IFS HRES in forecasting these extremes in the 0-5 day range. Bonavita (2024) argues that AI emulators trained to maximize RMSE over space and time may be better conceived as "forecast applications targeted at optimizing specific aspects of forecast performance, that is, medium-range mean squared/absolute errors over a range of atmospheric and near surface weather patterns" (Bonavita, 2024).

As AI emulator development continues, more focus is being placed on the forecast of extreme events. The use of ensembles, particularly huge ensembles, could improve the ability of these emulators to capture extreme events (Eyring et al., 2024; Mahesh, Collins, Bonev, Brenowitz, Cohen, Elms, et al., 2025). Along with increased ensemble size, some argue for focusing the training or post-processing of AI emulators on extremes (Olivetti & Messori, 2024a). Missing from this conversation is the understanding of how AI emulators currently learn extreme weather. Do they learn these events differently from more typical weather patterns? Is there a smoothing to the climatology that occurs over training that prevents the emulators from achieving high skill in forecasting extremes? We take a first look at these questions leveraging a spherical Fourier neural operator (SFNO), the architecture used in FourCastNet v2 (Bonev et al., 2023), and a collection of global extreme event cases including heatwaves, ARs, and TCs (Table 1).

| | **Event**,<br>Forecast Date | Description and Impact Snapshot |
|---|---|---|
| Heatwaves | **Pacific Northwest,**<br>2021-06-27 18Z | • Portland, Oregon record 116°F, Washington record 120°F<br>• In the top 6 most extreme heatwaves globally since 1960<br>• >800 deaths<br>(Loikith & Kalashnikov, 2023; Thompson et al., 2022; White et al., 2023) |
| | **Northern European,**<br>2022-06-26 18Z | • Tromsø, Norway above the Arctic circle was 86°F<br>• Two counties recorded highest temperature on June 29th<br>(Freedman, 2022; Meteorological Institute, 2022) |
| | **Antarctic,**<br>2022-03-18 12Z | • Widespread 30-40°C temperature anomalies<br>• Record high maximum March temperatures<br>• Most extensive melt event in the satellite era to occur outside of summer (December–February)<br>• Lead to the collapse of the Conger Ice Shelf<br>(Wille et al., 2024a, 2024b) |
| Atmospheric Rivers | **Midwest,**<br>2019-03-13 12Z | ● Associated bomb cyclone set a record for lowest pressure recorded over Colorado<br>● AR category 5 with IVT magnitudes >1,500 kg $m^{-1}$ $s^{-1}$<br>● Over 1,000 weather watch, warning and advisory statements<br>● Cancellation of ~1,400 flights at Denver International Airport<br>(National Weather Service, n.d.; Zou et al., 2025) |
| | **New Zealand,**<br>2019-03-25 12Z | ● Broke national rainfall record with 1086 mm (~43 inches) falling over 48 hours<br>● AR category 4<br>● Destroyed Waiho bridge<br>● Estimated $2.35 million in damages<br>(Prince et al., 2021) |
| | **Greenland,**<br>2022-03-14 12Z | ● Delivered 11.6 Gt of snowfall on March 14th<br>● Delayed summer melt by 11 days<br>● Offset Greenland's hydrological year net mass loss by 8%<br>(Bailey & Hubbard, 2025) |
| Tropical Cyclones | **Hurricane Ian,**<br>2022-09-29 0Z | ● Impacted Cuba and Florida<br>● Reached category 5, made landfall in Florida as category 4<br>● Maximum winds of more than 126 mph<br>● Storm surge up to 3-4.5 m, rainfall 13–17 inches per day<br>● More than 150 people died in the U.S.<br>● $112 billion in damages<br>(Bucci et al., 2023; Colby et al., 2024; Smith, 2020) |
| | **Cyclone Idai,**<br>2019-03-14 12Z | ● Impacted Mozambique, Zimbabwe, and Malawi<br>● Made landfall as a category 3 storm<br>● Winds above 120 mph<br>● >750 deaths - second deadliest cyclone in the southern hemisphere<br>● >715 Hectares of crops destroyed<br>(Mavhura & Aryal, 2022; UN Office of Coordination of Humanitarian Affairs, 2019; Warren, 2019) |
| | **Hurricane Dorian,**<br>2019-09-04 0Z | ● Greatest impact in the Bahamas<br>● Category 5 hurricanes with peak landfall winds 184 mph<br>● 22.84 inches of storm rainfall in Hope Town, Bahamas<br>● >200 people died<br>● 29,500 people left homeless and/or jobless<br>● $3.4 billion in damages in the Bahamas<br>(Avila et al., 2020) |

**Table 1:** Summary of extreme events selected for study.

## 2 Methods

The model and subsequent analysis uses ERA5 (Hersbach et al., 2020) at a 0.25 degree latitude-longitude resolution.

### 2.1 Extreme Events

Extreme events including heatwaves, ARs, and TCs are chosen from three years (2019, 2021, and 2022) of available data outside the training set. These events are selected to represent global impacts (see Table 1) and were selected prior to seeing any results. For analysis, we choose an available variable closely related to impact for each extreme weather type: 2 meter temperature (T2M) for heatwaves, total column water vapor (TCWV) for ARs, and mean sea level pressure (MSL) for TCs. We select an appropriate region of analysis based on the anomalies of each event (white area, Figures 1–3 a–c). For the heatwave events we limit our analysis to T2M over land.

We define three groups of timestamps: 1) Extreme events are selected timestamps of greatest impact for each event (see Table 1 for specific dates). 2) Climatology is the model's training set: 1979–2015 6-hourly ERA5. Climatology is used to quantify the extreme event anomaly. 3) Reference days for each extreme event include daily timestamps in 2019, 2021, and 2022 in the same month and hour as the extreme event. We remove four days before and after the extreme event from the reference days to define a distinct comparison group. The number of reference timestamps range from 81 to 84 days per event and are used to define forecast error over more typical weather days.

We quantify the anomaly of extreme events in two ways. First, we plot the variable anomaly compared to climatology over the region of analysis to demonstrate the spatial pattern of the event (Figures 2–4, a-c). Next, we calculate the absolute mean difference between climatology and each extreme event (stars in Figure S1) and reference timestamps (grey boxplots in Figure S1). This captures the anomaly magnitude compared to the reference days.

### 2.2 Model

We use a Spherical Fourier Neural Operator (SFNO) model, the same architecture used in AI2 Climate Emulator (ACE2) (Watt-Meyer et al., 2025) and NVIDIA's FourCastNet V2 (Bonev et al., 2023). We use the framework from the Mahesh et al. (2025) huge ensemble, with an SFNO scale factor of 3 and embedding dimension of 384. The model is built from the open source SFNO v0.1.0 in the modulus-makani Python repository ("NVIDIA/makani," September 26, 2023/2025). The SFNO model includes an encoder (a dense neural network that maps each latitude-longitude grid cell to a higher dimensional latent space), 8 SFNO blocks (spherical convolutions applied in the spectral domain followed by a dense neural network), and a decoder (a dense neural network that maps the latent space back into the physical space). SFNOs respect spherical geometry and retain physically plausible dynamics (Bonev et al., 2023). See Bonev (2023) Figure 3 for a schematic of this architecture and Mahesh et al. (2025) for a complete model description and discussion of hyperparameters, such as scale factor and embedding dimension.

Training data includes 6-hourly ERA5 reanalysis from 1979–2015 for 74 variables and 13 pressure levels for relevant variables (Mahesh, Collins, Bonev, Brenowitz, Cohen, Elms, et al., 2025). The cosine of the solar zenith angle, orography, and land-sea mask are included as non-prognostic inputs. The training of this SFNO model is identical to the training of a single ensemble member in Mahesh et al. (2025) with two additions: 1) After 70 epochs of training with a cosine annealing learning rate scheduler starting at 1e-3, we include a 20 epoch 2-step fine-tuning with a cosine annealing learning rate scheduler starting at 1e-4. 2) After each training epoch, model weights are saved to use in analysis.

The model is initiated with random weights and trained to minimize a mean-squared error between forecasts and ERA5 training data. Each epoch of training iterates through all training timesteps and the model is trained to forecast the next 6 hour timestep. All other details of training and hyperparameter selection are found in Mahesh et al. (2025). The model's overall best checkpoint evaluated on validation years 2016 and 2017 is checkpoint 70 without fine-tuning, and checkpoint 89 with fine-tuning. We reference both of these checkpoints when analyzing extreme forecast error.

**2.3 Forecast Runs and Analysis**

We choose a 5-day lead time to focus on synoptic-scale features that drive these extreme events. We complete a 5-day (20 timestep rollout) SFNO inference run for each extreme event and every reference timestamp using the model weights from each of the 90 training checkpoints. This results in 90 forecasts for each extreme event and 90 forecasts for each reference timestamp. The testing error over 90 checkpoints represents the evolution of the forecast error through training. Figure S2 displays an example of the progression of Midwest AR forecast through training.

We assess the regional pattern learned (Figures 1–3, d–f) using the latitude-weighted root mean square error (RMSE) between the model forecast and the ERA5 timesteps. We assess the regional extreme (Figures 1–3, d-f) using the absolute difference between the maximum or minimum variable value in the forecast versus the ERA5 timestamp.

We implement a gaussian smoothing (sigma = 2) on forecast error metrics over 90 checkpoints to capture the trends of learning and to identify the best checkpoint not attributable to stochastic noise. The smoothed best checkpoints (stars in Figure 1–3, d–i ) for extreme events do not necessarily represent the singular best forecast. Instead, they represent the point of training at which the model sustained the lowest error over multiple epochs.

## 3 Results, or a descriptive heading about the results

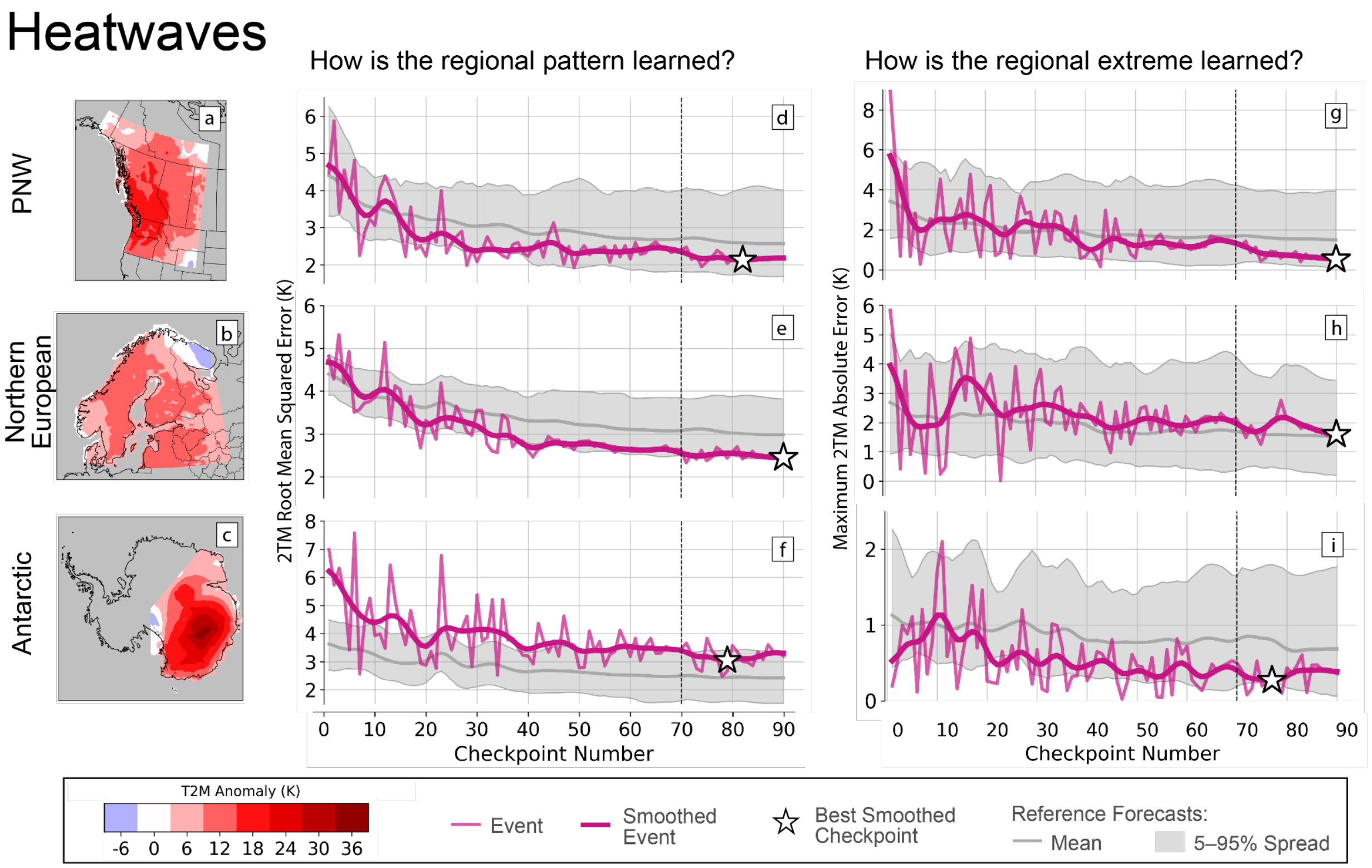

**Figure 1** Results for the PNW (a, d, g), Northern European (b, e, h) and Antarctic (c, f, i) Heatwaves. ERA5 T2M anomaly compared to 1979–2015 climatology (a–c) for the forecast date (Table 1) over the region of analysis (colored shading). Region of analysis forecast errors through 90 training checkpoints (d–i) for heatwave events (narrow pink lines), smoothed forecasts through training (thick pink lines), smoothed reference forecast mean (grey line), and smoothed reference forecast 5–95% spread (grey shading). White stars indicate the best smoothed checkpoint, and the vertical dashed line marks the start of 2-step fine tuning during training. The regional pattern learned is quantified by the T2M RMSE (e–f) while the regional extreme learned is quantified by the maximum T2M absolute error (g–i).

### 3.1 Heatwaves

The three heatwaves analyzed include two record-shattering events in the Pacific Northwest and Antarctica, and one less extreme heatwave in Northern Europe (Table 1). The PNW and Antarctic heatwaves are both "gray swan" events – the model did not see any similarly extreme regional heatwave events in the training dataset (Thompson et al., 2022; Wille et al., 2024a). The PNW, Northern European, and Antarctic T2M anomalies reach 19, 15, and 38 Kelvin above 1980–2015 monthly climatology in ERA5, respectively (Figure 1 a–c). The T2M anomalies of all three heatwaves surpass those of any reference timestamp, but the PNW and Antarctic heatwaves fall significantly outside the distribution: 3.7 and 5.4 standard deviations from the reference forecast mean T2M anomaly, respectively (Figure S1a).

Despite representing anomalous T2M environments compared to reference days, the SFNO forecasts these heatwaves with comparable error to reference forecasts. At the model's best checkpoint before finetuning (checkpoint 70) and the model's best fine-tuning checkpoint (checkpoint 89), the heatwave forecast error falls in the 5–95% range of reference forecasts, often outperforming the mean (Figure 1).

The similarities between heatwave forecasts and reference forecasts is not limited to the model's best checkpoints. With the exception of the Antarctic heatwave regional pattern, the smoothed extreme event forecasts are indistinguishable from the smoothed reference forecasts, falling within the 5–95% spread throughout training. This result holds true for unsmoothed forecasts as well (Figure S3). As one might expect, reference forecasts get better over training, and the smoothed heatwave forecasts behave similarly, with the best smoothed checkpoint for both regional pattern and regional extreme falling in the last 15 epochs (Figure 1).

Some recent work has highlighted the difficulty AI emulators face in forecasting out-of-training-distribution grey swan events, finding evidence that AI emulators are better able to learn extreme events via translocation (learning from similar events in a different region) than they are able to extrapolate from weaker events in the same region (Sun, Hassanzadeh, Shaw, et al., 2025; Sun, Hassanzadeh, Zand, et al., 2025). We find that for these three heatwaves, the two grey swan events (PNW and Antarctic) are not substantially different through training compared to the less extreme Northern European heatwave. This may indicate that the SFNO is learning via translocation for these events.

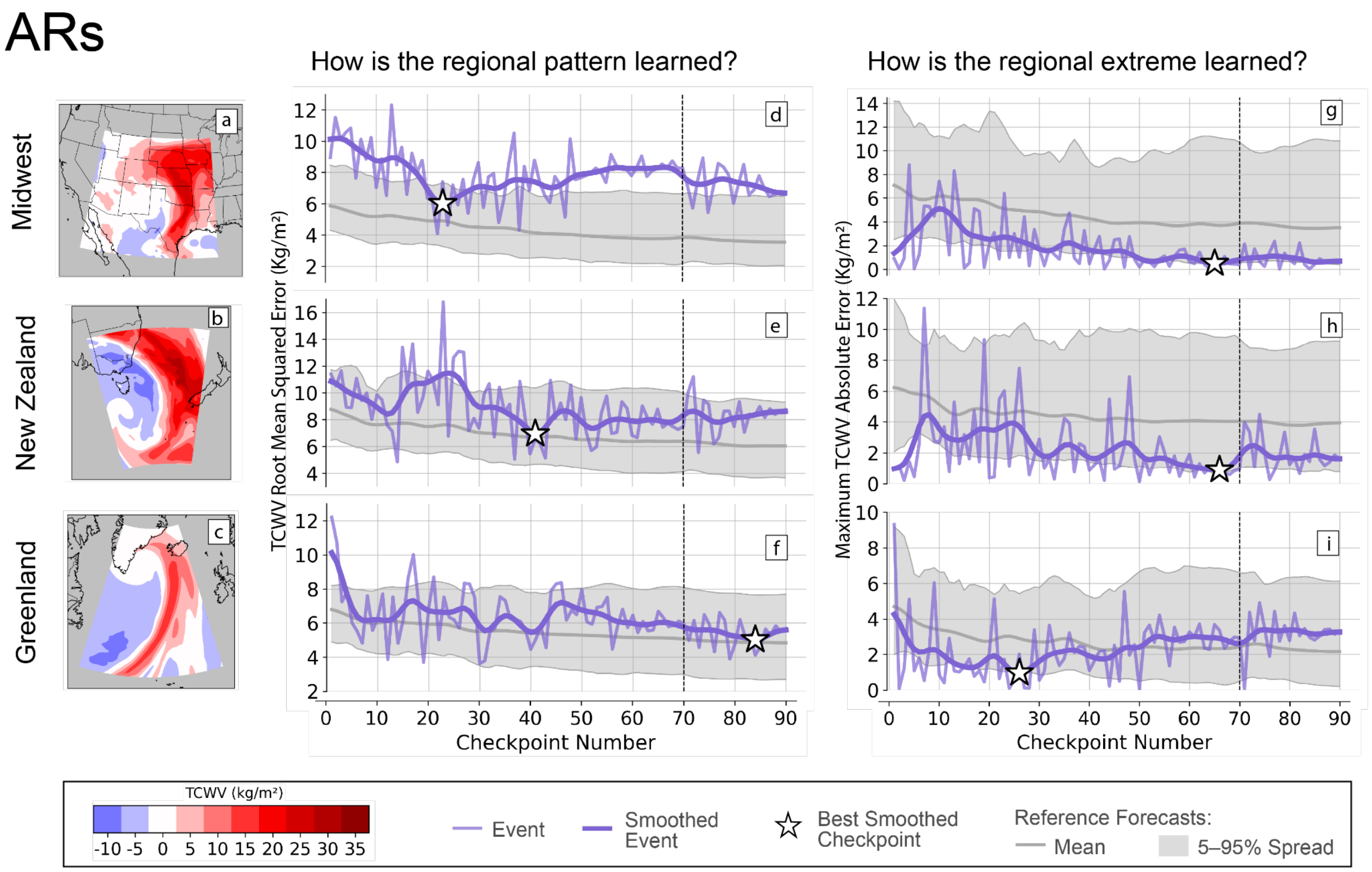

**Figure 2** Results for AR events. As in Figure 1, but for the Midwest (a, d, e), New Zealand (b, e, h), and Greenland (c, f, i) ARs using TCWV RMSE and maximum TCWV absolute error.

### 3.2 Atmospheric Rivers

We assess three AR events all associated with significant impacts (Table 1) but ranging in extremity as defined by their anomalous TCWV. The Midwest AR has a greater TCWV anomaly than any reference forecast, while the New Zealand AR falls in the 97th percentile of reference forecast TCWV anomalies and the Greenland AR only falls in the 83rd. The maximum TCWV anomalies compared to climatology are also more extreme for the Midwest AR (28 kg/m$^2$) and the New Zealand AR (33 kg/m$^2$) compared to the Greenland AR (16 kg/m$^2$) (Figure 2 a–c).

At the model's best checkpoint before fine tuning (checkpoint 70), the regional pattern of all three ARs are forecasted with larger error than the average reference forecasts (Figure 2, d–f). The same is true for the model's best fine-tuning checkpoint. However, the checkpoint 70 forecast captures maximum TCWV with similar or lower error than reference forecasts (Figure 2, d–f). This is true for the best fine-tuning checkpoint as well, except for the Greenland AR forecast (Figure 2i). This suggests that for the more anomalous ARs, the SFNO is better able to capture the extreme moisture than the exact placement of the AR (e.g. Figure S2).

The progression of the AR forecasts through training is starkly different from the heatwaves. While the majority of AR smoothed forecasts lie within the 5–95% spread of smooth reference forecasts throughout training, the AR forecasts do not steadily improve through training. Instead, the Midwest AR regional pattern, the New Zealand AR regional pattern, and the Greenland AR regional extreme are best forecasted in the first half of training (checkpoints 22, 40, and 25 respectively). For all three AR events, the best smoothed checkpoint forecast falls well within the 5–95% spread of the reference forecasts, after which, the forecast degrades with more training. This is most obvious in the Midwest AR regional pattern forecasts (see also Figure S2 for the spatial pattern).

### 3.3 Tropical Cyclones

The regional MSL anomaly of Hurricane Ian, TC Idai, and Hurricane Dorian fall in the 90th, 88th, and 89th percentile of reference days, respectively (Figure S1). We note an obvious difference in forecast error for TCs compared to ARs and heatwaves: TC regional patterns and minimum MSL values are poorly forecasted compared to reference days. This is consistent with previous research identigying a low bias in AI emulator cyclone intensity (DeMaria et al., 2024, 2025). At the SFNO best checkpoint 70 and the best fine tuning checkpoint 89, both the regional pattern and minimum MSL lie far outside the 5–95% spread of reference forecasts (Figure 3). Throughout training, most checkpoints underperform compared to the 5–95% spread of reference forecasts. We note that, unlike heatwaves and ARs, even the best smoothed TC checkpoints underperform compared to the 5–95% spread of reference forecasts. Similar to a pattern we saw in some AR forecasts, the TC forecasts degrade with further training after the best smoothed checkpoint early in training. Some of these trends are marginal (e.g. regional pattern smooth forecast for Cyclone Idai) while others result in a 70th and 89th checkpoint RMSE ~6x higher than the checkpoint at the best smoothed checkpoint epoch (regional extreme forecast for Hurricane Ian).

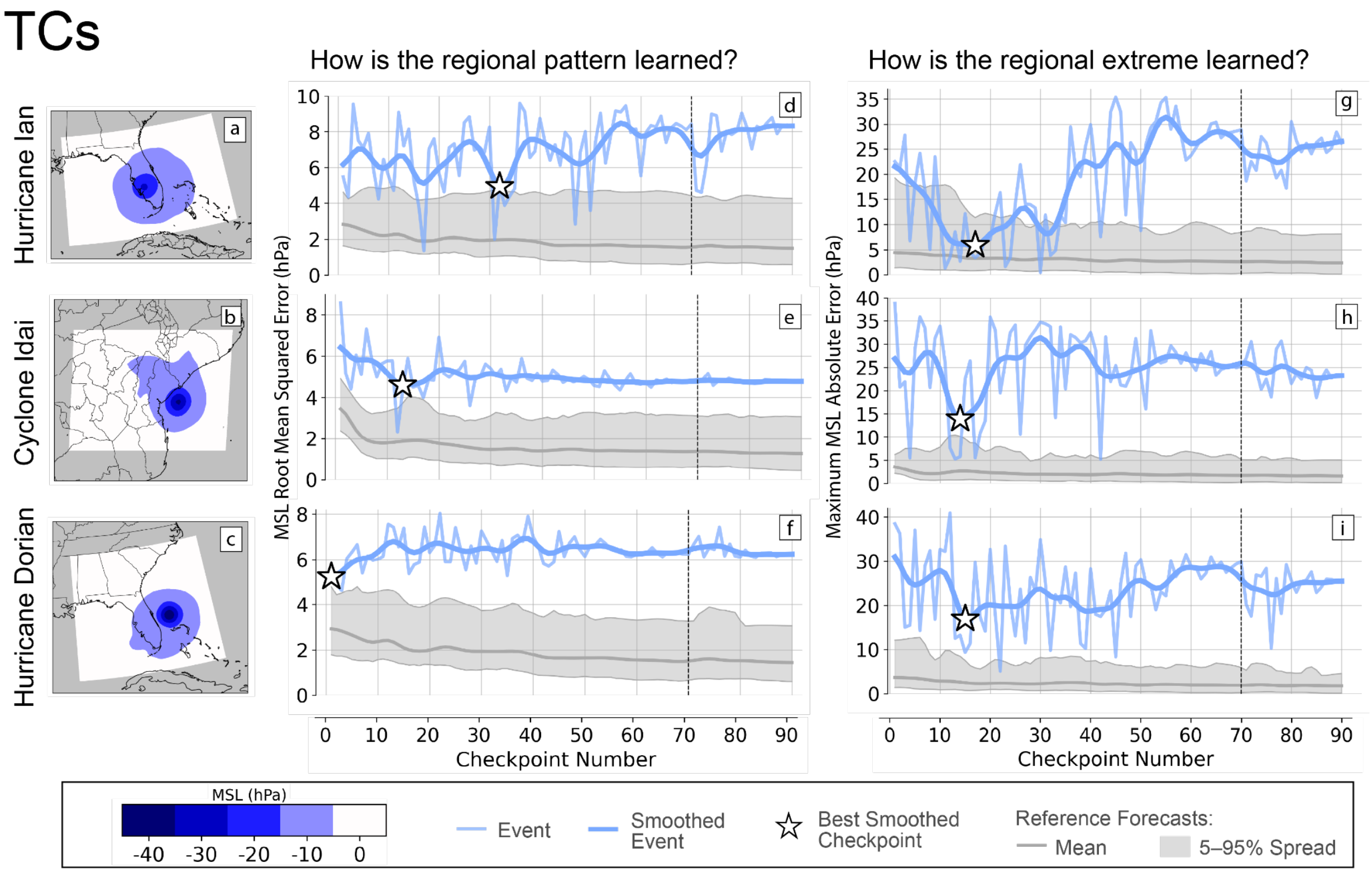

**Figure 3** Results for TC events. As in Figure 1, but for Hurricane Ian (a, d, e), Cyclone Idai (b, e, h), and Hurricane Dorian (c, f, i) using MSL RMSE and minimum MSL absolute error.

## 4 Discussion

We present an initial investigation of how an AI emulator learns extreme weather forecasts during training and how this differs from its learning of more typical weather days. Despite the limited scope of this study (a single deterministic AI emulator, nine extreme events, three years of reference forecasts, 2 events from EMA models), this work meaningfully contributes to a growing body of research that assesses the ability of various AI emulators to skillfully forecast extreme events (Davis et al., n.d.; Olivetti & Messori, 2024b, 2024a; Pasche et al., 2025).

Our results support Olivetti & Messori (2024a) in calling for more focus on extreme weather in the development of AI emulators. Recent research has answered this call by creating huge ensembles (Mahesh, Collins, Bonev, Brenowitz, Cohen, Elms, et al., 2025; Mahesh, Collins, Bonev, Brenowitz, Cohen, Harrington, et al., 2025) and designing training frameworks to better target extreme events (Olivetti & Messori, 2024a). Adding to this conversation, our results suggest that AI emulators may be learning information about extreme events during training that is lost by the model's best training epoch.

To improve extreme weather forecasts based on our results, one may be tempted to implement early stopping during training based on forecast skill for a particular extreme weather type. However, the best training checkpoint for each weather type (heatwave, AR, and TC) varies by event. Additionally, the reference forecasts at these training checkpoints are generally worse than the model's best checkpoint. Thus early stopping is likely not a viable method for harnessing information from earlier training epochs to improve extreme event forecasts.

Exponential moving average (EMA) of weights in deep learning is a technique in which the model weights are smoothed through training. A decay rate parameter determines what percentage of the new weight is determined by the previous weights. For image classification, EMA models with early stopping are shown to outperform stochastic gradient descent models and consistently improve generalization (Morales-Brotons et al., 2024). The AI2 Climate Emulator (ACE) saw up to a 15% improvement in time-mean RMSE with EMA (Watt-Meyer et al., 2023), and EMA is included in ACE2 (Watt-Meyer et al., 2025). However, EMA has not been explored for the explicit purpose of improving extreme weather forecasts. We hypothesize that implementing EMA on the SFNO model might improve extreme weather forecasts by retaining some information from earlier checkpoints.

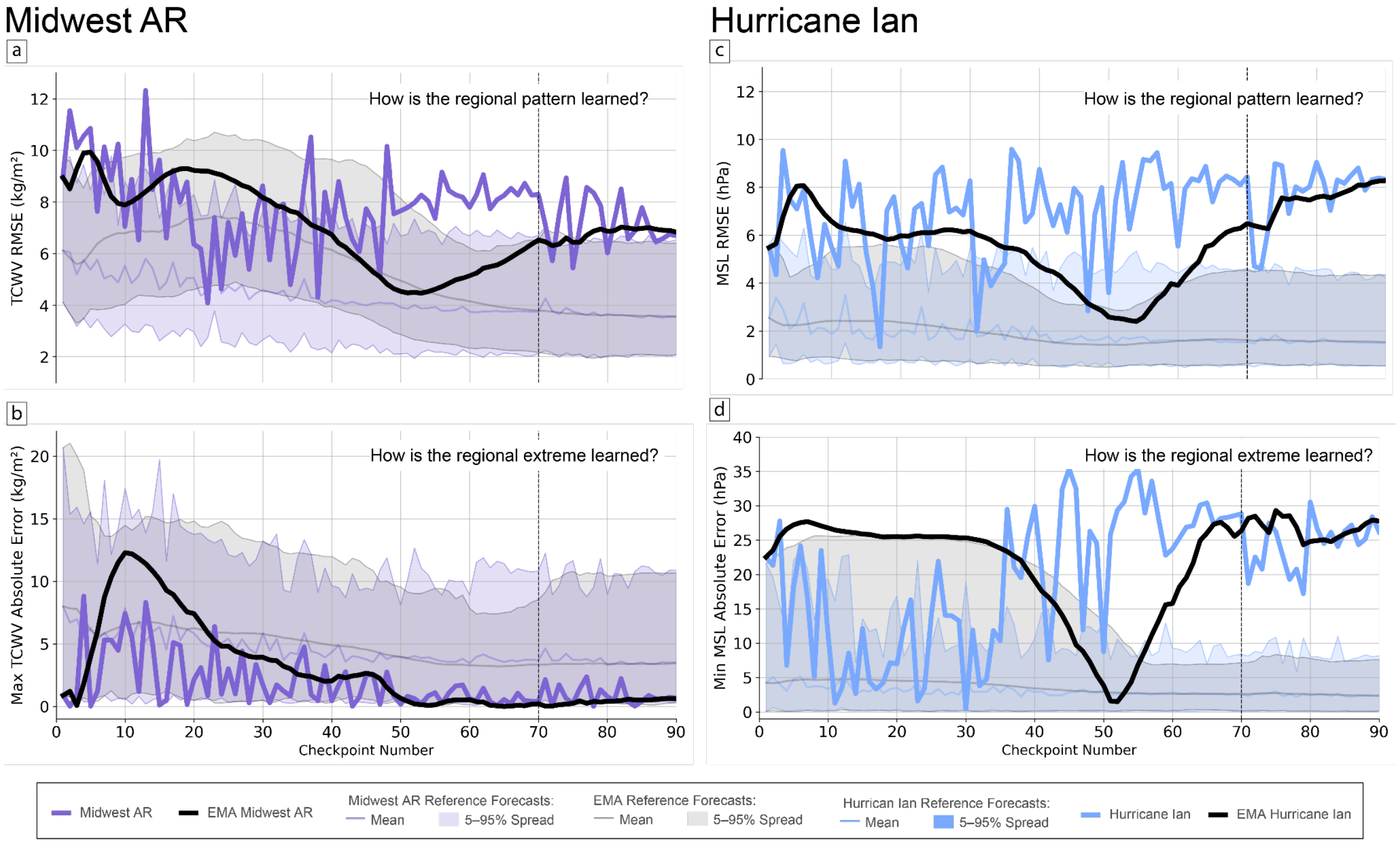

**Figure 4** Results from EMA model compared to SFNO model for the Midwest AR (a,b) and Hurricane Ian (c, d). Region of analysis forecast error metrics through 90 training checkpoints (a–d) for extreme events from SFNO model (thick purple and blue lines), extreme events from EMA model (thick black lines), SFNO model reference forecast mean (purple and blue lines), EMA model reference forecast mean (grey lines), SFNO reference forecast 5–95% spread (purple and blue shading), and EMA reference forecast 5–95% spread (grey shading). The vertical dashed line marks the start of 2-step fine tuning during training. The regional pattern learned is quantified by the TCWV (a) and MSL (c) RMSE while the regional extreme learned is quantified by the maximum TCWV (b) and minimum MSL (d) absolute error.

We provide a first test of this hypothesis with two of our extreme events (Midwest AR and Hurricane Ian) and their associated reference forecasts. We use a relatively small decay value of 0.9, meaning that each checkpoint of training is 90% information from the previous checkpoints and 10% new weights. As expected, the EMA forecasts are smoother than those without EMA implementation (Figure 4). For all checkpoints between 50–70, the EMA model outperforms the original SFNO in forecasting the Midwest AR and Hurricane Ian. At Checkpoint 70, the EMA model forecasts the extreme events with smaller errors while maintaining similar errors in the reference forecasts. Additionally, we see a drastic decrease in error for extreme forecasts around checkpoint 52 with similar reference forecast error in all metrics except for Midwest AR regional pattern reference forecasts that underperforms compared to the original SFNO without EMA. The EMA performance in fine tuning epochs 71–90 degrades for extreme events and levels off for reference forecasts, consistent with the finding that EMA models have the best performance with large learning rates (Morales-Brotons et al., 2024). With these two examples, we provide evidence for the potential benefit of EMA models for forecasting of extreme events in AI emulators while maintaining time-averaged error. We encourage a more comprehensive investigation of these potential benefits with a larger sample of extreme events, especially because the addition of an EMA model to deep learning models during training is relatively low-effort and low-cost (Morales-Brotons et al., 2024).

Our results reveal some important insights into medium-range forecasting of extreme weather events in AI emulators: 1) AI emulators may be learning information about forecasting extreme events like ARs and TCs earlier in training and losing that information by the end of training when the time-averaged error is minimized. 2) Given this result, we propose more focus on novel training methods for AI weather emulators that harness learning of extreme weather while maintaining forecast skill for more typical weather days. We provide evidence of improvement with one such method.

## Supporting Information

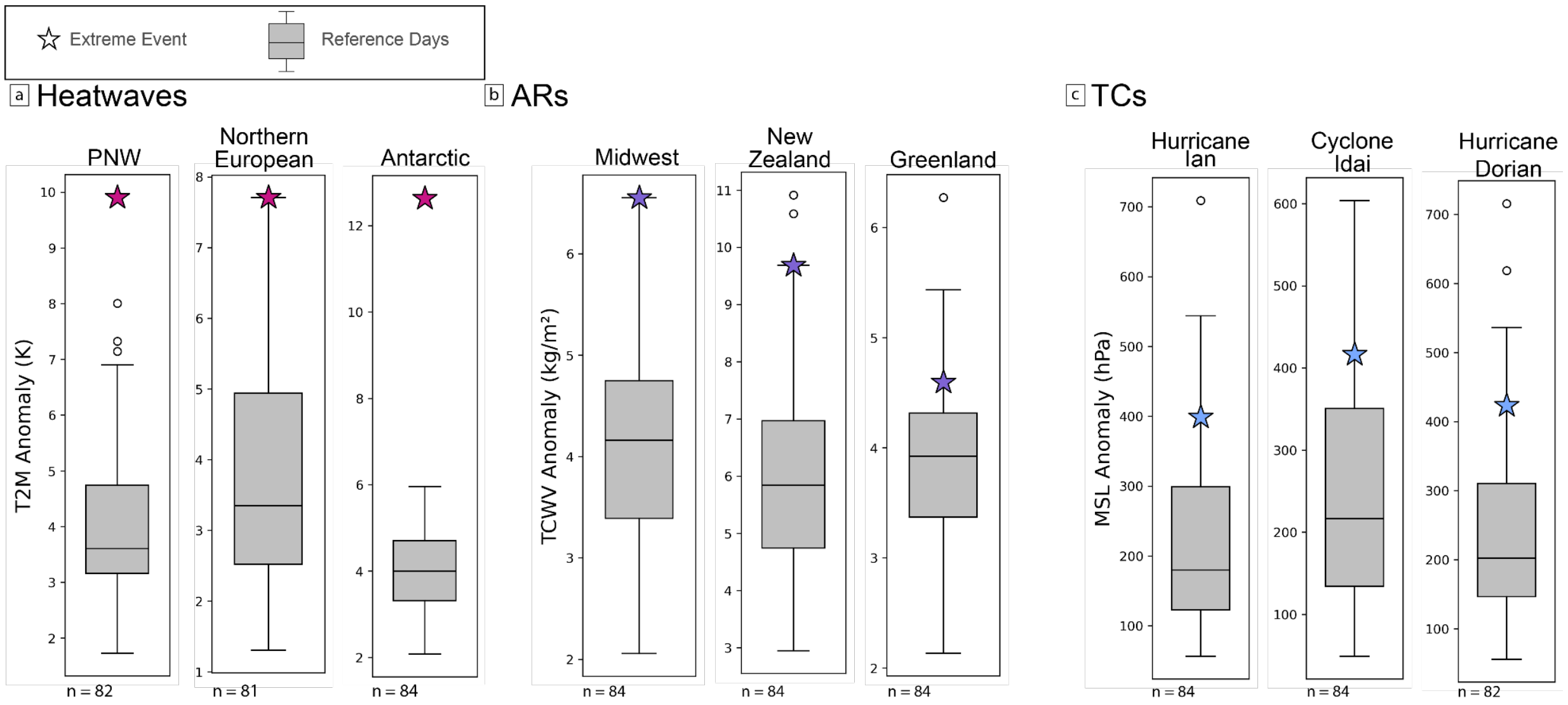

**Figure S1.** Regional anomaly of extreme events. Region of analysis T2M (a), TCWV (b), and MSL (c) mean anomalies compared to 1979–2015 monthly climatology for extreme events (stars) and reference timestamps (grey boxplots). The number of reference timestamps for extreme event is shown below each plot.

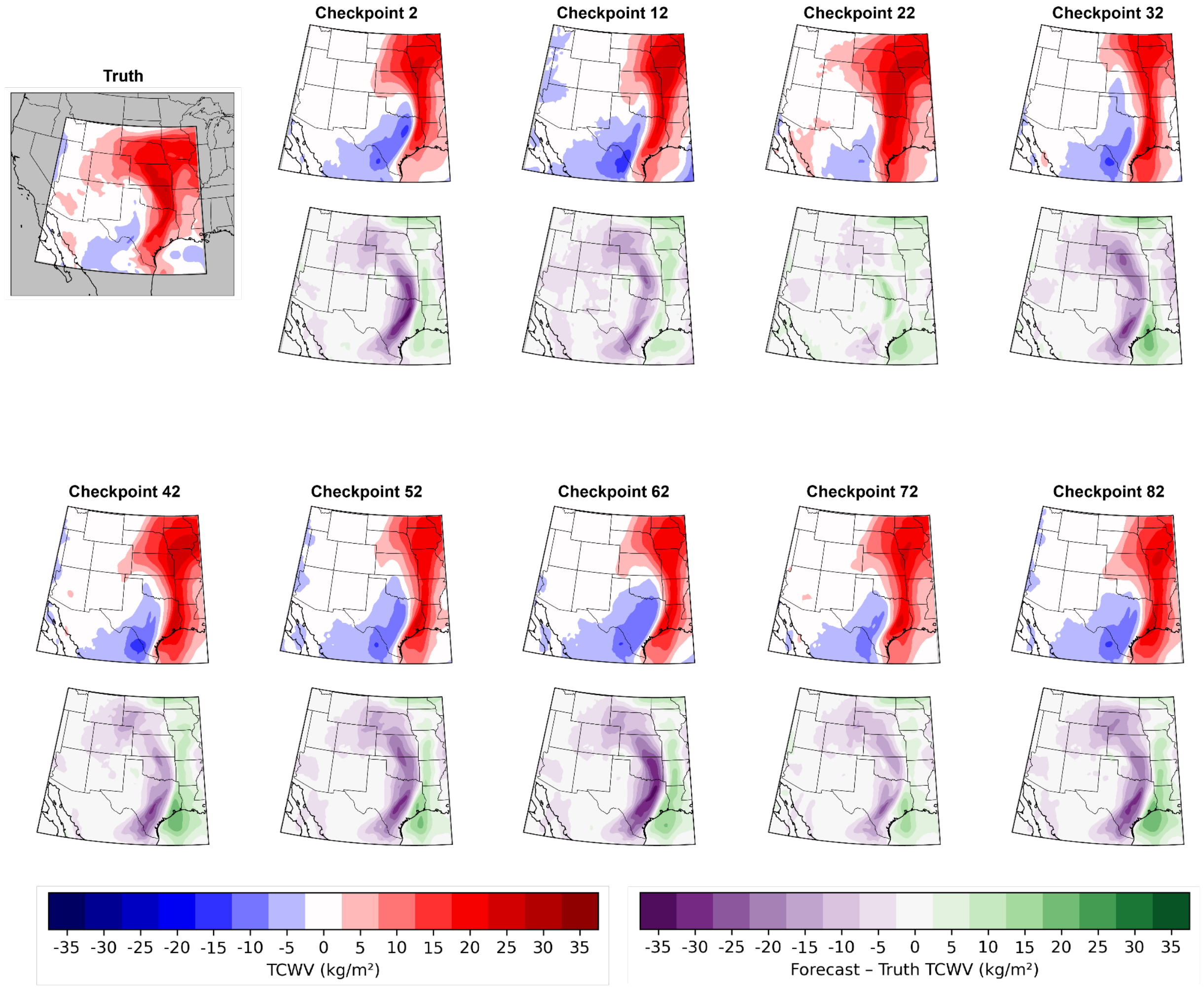

**Figure S2.** Training evolution for the Midwest AR. ERA5 TCWV on March 13, 2019 at 12Z in the region of analysis (top left). SFNO 5 day forecasts for March 13, 2019 at 12Z from every 10th training checkpoint (red and blue shading) and difference between ERA TCWV and forecast at each checkpoint (purple and green shading).

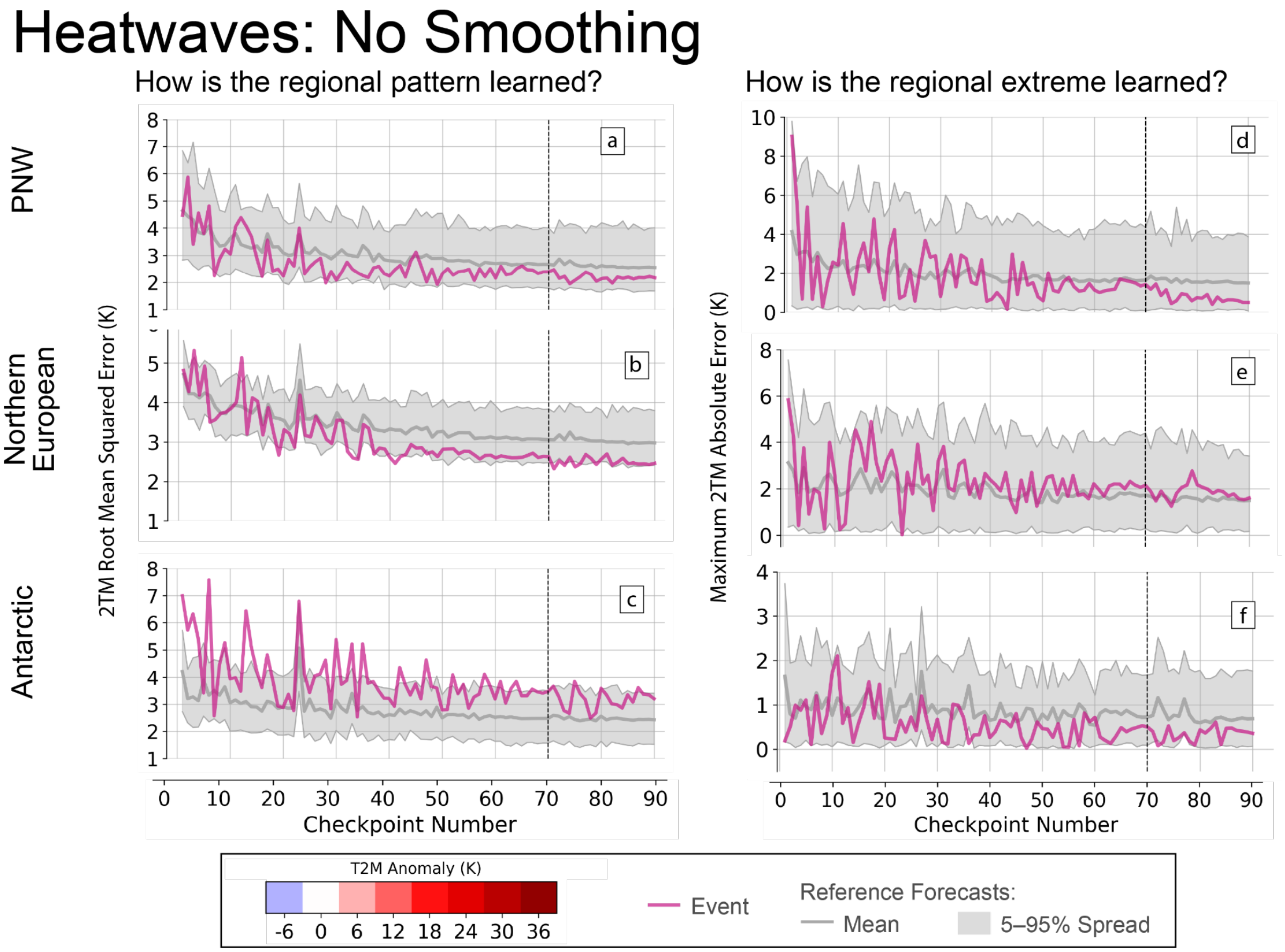

**Figure S3.** Results for the PNW (a, d, g), Northern European (b, e, h) and Antarctic (c, f, i) Heatwaves. ERA5 T2M anomaly compared to 1979–2015 climatology (a–c) for the forecast date (Table 1) over the region of analysis (colored shading). Region of analysis forecast errors through 90 training checkpoints (d–i) for heatwave events (pink lines), reference forecast mean (grey line), and reference forecast 5–95% spread (grey shading). White stars indicate the best smoothed checkpoint, and the vertical dashed line marks the start of 2-step fine tuning during training. The regional pattern learned is quantified by the T2M RMSE (e–f) while the regional extreme learned is quantified by the maximum T2M absolute error (g–i).

**Acknowledgments**

This research was supported by Quadrature Climate Foundation Grant 01-21-000338 and Founder's Pledge award 240905. This research was supported by the Director, Office of Science, Office of Biological and Environmental Research of the US Department of Energy under contract no. DE-AC02-05CH11231 and by the Regional and Global Model Analysis Program area within the Earth and Environmental Systems Modeling Program. The research used resources of the National Energy Research Scientific Computing Center (NERSC), also supported by the Office of Science of the US Department of Energy, under contract no. DE-AC02-05CH11231. The computation for this paper was supported in part by the DOE Advanced Scientific Computing Research (ASCR) Leadership Computing Challenge (ALCC) 2023–2024 award "Huge Ensembles of Weather Extremes using the Fourier Forecasting Neural Network" to William Collins (LBNL) and ALCC 2024–2025 award "Huge Ensembles of Weather Extremes using the Fourier Forecasting Neural Network" to William Collins (LBNL).

**Open Research**

Code and information for SFNO inference runs is available at https://nvidia.github.io/earth2studio/ ("NVIDIA/earth2studio: Open-source deep-learning framework for exploring, building and deploying AI weather/climate workflows.," n.d.) under the Apache-2.0 license. For training epoch model weights, contact the author.

**Conflict of Interest Disclosure**

The authors declare there are no conflicts of interest for this manuscript.